\documentclass[aps,amsmath,amssymb,prc,groupedaddress,nofootinbib]{revtex4-2}
\usepackage{bm}
\usepackage{braket}
\usepackage{color}
\usepackage{dcolumn}
\usepackage{graphicx}
\usepackage{hyperref}
\usepackage{multirow}

\def\ColorA{black}
\def\ColorB{black}
\def\ColorC{black}
\def\ColorD{black}
\def\ColorE{black}
\newcommand{\RevA}[1]{{\color{\ColorA}{#1}}}
\newcommand{\RevB}[1]{{\color{\ColorB}{#1}}}
\newcommand{\RevC}[1]{{\color{\ColorC}{#1}}}
\newcommand{\RevD}[1]{{\color{\ColorD}{#1}}}
\newcommand{\RevE}[1]{{\color{\ColorE}{#1}}}

\begin{document}

\title{Equation of state for hyperonic neutron-star matter in SU(3) flavor symmetry}
\author{Tsuyoshi Miyatsu}
\email[]{tsuyoshi.miyatsu@ssu.ac.kr}
\affiliation{Department of Physics and OMEG Institute, Soongsil University, Seoul 06978, Republic of Korea}
\author{Myung-Ki Cheoun}
\email[]{cheoun@ssu.ac.kr}
\affiliation{Department of Physics and OMEG Institute, Soongsil University, Seoul 06978, Republic of Korea}
\author{Kyungsik Kim}
\email[]{kyungsik@kau.ac.kr}
\affiliation{School of Liberal Arts and Sciences, Korea Aerospace University, Goyang 10540, Republic of Korea}
\author{Koichi Saito}
\email[]{koichi.saito@rs.tus.ac.jp}
\affiliation{Department of Physics and Astronomy, Tokyo University of Science, Noda 278-8510, Japan}
\date{\today}

\begin{abstract}
  Using a relativistic mean-field model calibrated to finite-nucleus observables and bulk properties of dense nuclear matter, we investigate hyperonic neutron-star matter within an SU(3) flavor-symmetry scheme.
  To retain SU(6)-based couplings within SU(3) flavor symmetry, we introduce a quartic $\phi$ self-interaction and $\phi$-$\rho$ mixing.
  We demonstrate the roles of $\alpha_{v}$ ($F/(F+D)$ ratio), $\theta_{v}$ (mixing angle), and $z_{v}$ (singlet-to-octet coupling ratio) in SU(3)-invariant vector-meson couplings.
  It is found that $z_{v}$ predominantly controls the maximum mass of a neutron star, and $2M_{\odot}$ neutron stars can be supported for $z_{v}\le0.15$.
  The $\alpha_{v}$ also helps sustain large masses, whereas $\theta_{v}$ has a smaller effect on neutron-star properties.
  This SU(3) framework reconciles nuclear and astrophysical constraints, and offers a plausible resolution of the \textit{hyperon puzzle}.
\end{abstract}

\maketitle


\section{Introduction}
\label{sec:introduction}

Recent multimessenger observations have significantly advanced our understanding of the dense matter equation of state (EoS).
In particular, precise measurements of neutron-star radii from the \RevA{Neutron star Interior Composition ExploreR (NICER)} mission~\citep{Vinciguerra:2023qxq,Choudhury:2024xbk,Salmi:2024aum,Salmi:2024bss,Mauviard:2025dmd}, together with the constraints on the tidal deformability from the binary neutron-star merger event, GW170817~\citep{LIGOScientific:2017vwq,LIGOScientific:2018cki,LIGOScientific:2018hze}, have placed stringent limits on the properties of dense matter at supranuclear densities.
These astrophysical constraints, combined with the discovery of massive pulsars with masses around or above $2M_{\odot}$~\citep{Demorest:2010bx,Antoniadis:2013pzd,NANOGrav:2017wvv,NANOGrav:2019jur}, strongly influence the theoretical modeling of neutron-star matter and provide important clues to its microscopic composition.

The possible appearance of hyperons in the core of a neutron star is a natural outcome of the Pauli exclusion principle at high densities, where the conversion of high-momentum nucleons into hyperons becomes energetically favorable.
However, their onset generally softens the nuclear EoS, leading to a substantial reduction in the maximum mass of a neutron star, $M_{\textrm{max}}$.
This discrepancy between theoretical predictions and astrophysical observations is widely known as the \textit{hyperon puzzle}.

A variety of approaches has been proposed to address this problem.
In nonrelativistic frameworks, repulsive three-body forces among baryons have been introduced to provide additional stiffness at high densities, as demonstrated in variational calculations and Brueckner-Hartree-Fock (BHF) studies~\citep{Schulze:2006vw,Vidana:2010ip,Togashi:2016fky,Lonardoni:2014bwa}.
In relativistic calculations, several extensions have been explored, using relativistic mean-field (RMF) approaches with the additional strange mesons~\citep{Schaffner:1993qj,Schaffner:1995th,Sulaksono:2012ny,Tu:2021sxx} and/or explicit three-baryon couplings~\citep{Tsubakihara:2012ic,Muto:2021jms}, relativistic Hartree-Fock (RHF) models including Fock terms and tensor couplings~\citep{Miyatsu:2011bc,Katayama:2012ge,Miyatsu:2015kwa,Li:2018jvz,Li:2018qaw}, and Dirac-Brueckner-Hartree-Fock (DBHF) calculations~\citep{Sammarruca:2009wn,Katayama:2013zya,Katayama:2015dga}.


\RevE{The RMF model is a covariant energy-density functional framework that effectively describes baryonic interactions in terms of meson exchange~\citep{Ring:1996qi,Vretenar:2005zz,Meng:2005jv}, but it does not explicitly include quark and gluon degrees of freedom originating from microscopic dynamics based on \RevA{quantum chromodynamics (QCD)}.
  Therefore, it should be regarded as a phenomenological realization of the strong interaction, complementary to QCD-motivated approaches such as the lattice QCD~\citep{Aoki:2005vt,Beane:2010em,Kronfeld:2012uk}, chiral effective field theory ($\chi$EFT)~\citep{Epelbaum:2008ga,Drischler:2021kxf}, Dyson-Schwinger equation formalism~\citep{Roberts:1994dr,Roberts:2000aa}, Nambu--Jona-Lasinio (NJL) model~\citep{Klevansky:1992qe,Hatsuda:1994pi,Buballa:2003qv}, quark-meson coupling (QMC) model~\citep{Guichon:1987jp,Saito:1994ki,Guichon:1995ue,Saito:2005rv,Nagai:2008ai}, and Quarkyonic matter~\citep{Fujimoto:2023mzy,Fujimoto:2024doc,Kojo:2025vcq}.}

One of the most practical and effective approaches proposed to resolve the \textit{hyperon puzzle} in the RMF models is the adoption of SU(3) flavor symmetry for determining the vector-meson couplings to the baryon octet~\citep{Weissenborn:2011ut,Miyatsu:2013yta,Lopes:2013cpa,Oertel:2014qza,Spinella:2018dab,Fu:2022eeb,Lopes:2022vjx}.
This approach enables the tuning of model parameters with empirical hypernuclear data and hyperon potentials in nuclear matter, introducing an additional repulsive force via the $\phi$ meson.
Such flexibility can delay the onset of hyperons and maintain a sufficiently stiff EoS, which can support $2M_{\odot}$ neutron stars.
However, only a few studies have attempted to simultaneously account for both results from nuclear experiments and astrophysical observations.
In particular, the nuclear EoS at high densities often shows a substantial mismatch from the constraints obtained in heavy-ion collisions~\citep{Danielewicz:2002pu,Fuchs:2005zg,Lynch:2009vc}.

The effective nucleon mass, $M_{N}^{\ast}$, plays a crucial role in determining the global properties of neutron stars.
As discussed in \citet{Choi:2020eun} and \citet{Li:2024tpr}, a lower $M_{N}^{\ast}$ generally leads to a stiffer EoS, thereby increasing $M_{\textrm{max}}$ to values exceeding $2M_{\odot}$.
However, this reduction in $M_{N}^{\ast}$ also tends to enhance the dimensionless tidal deformability, $\Lambda$, making it challenging to reproduce the constraints from GW170817~\citep{LIGOScientific:2018cki,LIGOScientific:2018hze}.
This dual sensitivity of the EoS to $M_{N}^{\ast}$ highlights its pivotal role in balancing the competing requirements from both $M_{\textrm{max}}$ and $\Lambda$ observations.
Moreover, recent model-independent analysis suggests a much smaller $\Lambda$ than most current estimates~\citep{Huang:2025mrd}.
Such a stringent constraint is particularly difficult to accommodate in the standard RMF-based EoSs, posing a serious challenge to developing a unified description of dense nuclear matter consistent with astrophysical observations.

In the present study, we extend our previous work in \citet{Miyatsu:2013yta}.
We first build the RMF model that satisfies neutron-star radii and $\Lambda$ from NICER and GW170817 as well as the characteristics of finite nuclei in SU(6) spin-flavor symmetry.
The SU(3) flavor-symmetry approach, incorporating the effects of strange mesons, is then applied to the calibrated RMF parameters to examine how $M_{\textrm{max}}$ changes.
We aim to clarify the role of modified vector-meson couplings in balancing the stiffness of the hyperonic neutron-star EoS under the combined constraints from both astrophysical and terrestrial data.

This paper is organized as follows.
Section~\ref{sec:Lagrangian} reviews the extended RMF model based on Quantum Hadrodynamics~\citep{Serot:1984ey}.
The SU(3) extension of the vector-meson couplings is described in Sec.~\ref{sec:SU3}.
Section~\ref{sec:model} outlines the model construction and the determination of parameter sets consistent with both nuclear experiments and neutron-star observations.
The properties of neutron stars in SU(3) flavor symmetry are presented in Sec.~\ref{sec:results}.
Finally, Sec.~\ref{sec:summary} summarizes the main results and conclusions.

\section{Lagrangian density in SU(3) flavor symmetry}
\label{sec:Lagrangian}

We extend the standard Lagrangian density in the RMF approximation to include not only the $\sigma$, $\omega^{\mu}$, $\bm{\delta}$, and $\bm{\rho}^{\mu}$ mesons but also the strange mesons, namely the isoscalar, Lorentz-scalar ($\sigma^{\ast}$) and Lorentz-vector ($\phi^{\mu}$) mesons.
Since charge neutrality and $\beta$ equilibrium conditions are imposed in neutron-star matter, leptons are also included in the system.
The total Lagrangian density is, thus, chosen to be~\citep{Miyatsu:2013yta,Choi:2020eun,Miyatsu:2023lki,Miyatsu:2024ioc}
\begin{align}
  \mathcal{L}
  & = \sum_{B=N,\,\Lambda,\,\Sigma,\,\Xi} \bar{\psi}_{B} \left[ i\gamma_{\mu}\partial^{\mu} - M_{B}^{\ast}(\sigma,\bm{\delta},\sigma^{\ast})
    - g_{\omega B}\gamma_{\mu}\omega^{\mu} - g_{\rho B}\gamma_{\mu}\bm{\rho}^{\mu}\cdot\bm{I}_{B} - g_{\phi B}\gamma_{\mu}\phi^{\mu} \right] \psi_{B}
    \nonumber \\
  & + \frac{1}{2}\left(\partial_{\mu}\sigma\partial^{\mu}\sigma-m_{\sigma}^{2}\sigma^{2}\right)
    + \frac{1}{2}m_{\omega}^{2}\omega_{\mu}\omega^{\mu} - \frac{1}{4}W_{\mu\nu}W^{\mu\nu}
    \nonumber \\
  & + \frac{1}{2}\left(\partial_{\mu}\bm{\delta}\cdot\partial^{\mu}\bm{\delta} - m_{\delta}^{2}\bm{\delta}\cdot\bm{\delta}\right)
    + \frac{1}{2}m_{\rho}^{2}\bm{\rho}_{\mu}\cdot\bm{\rho}^{\mu} - \frac{1}{4}\bm{R}_{\mu\nu}\cdot\bm{R}^{\mu\nu}
    \nonumber \\
  & + \frac{1}{2}\left(\partial_{\mu}\sigma^{\ast}\partial^{\mu}\sigma^{\ast}-m_{\sigma^{\ast}}^{2}\sigma^{\ast2}\right)
    + \frac{1}{2}m_{\phi}^{2}\phi_{\mu}\phi^{\mu} - \frac{1}{4}P_{\mu\nu}P^{\mu\nu}
    \nonumber \\
  & - U_{\textrm{NL}}(\sigma,\omega^{\mu},\bm{\rho}^{\mu},\phi^{\mu})
    + \sum_{\ell=e,\,\mu} \bar{\psi}_{\ell} \left( i\gamma_{\mu}\partial^{\mu} - m_{\ell}\right) \psi_{\ell},
    \label{eq:Lagrangian}
\end{align}
where $\psi_{B}$ and $\psi_{\ell}$ denote baryon ($B$) and lepton ($\ell$) fields, respectively, and the sum runs $B\in\{N,\,\Lambda,\,\Sigma,\,\Xi\}$~\footnote{
  The baryon fields are expressed as 
  $
  \psi_{N} =
  \begin{pmatrix}
    p \\
    n \\
  \end{pmatrix}
  $,
  $
  \psi_{\Lambda} =
  \begin{pmatrix}
    \Lambda \\
  \end{pmatrix}
  $,
  $
  \psi_{\Sigma} =
  \begin{pmatrix}
    -\Sigma^{+} \\
    \Sigma^{0}  \\
    \Sigma^{-}  \\
  \end{pmatrix}
  $, and
  $
  \psi_{\Xi} =
  \begin{pmatrix}
    -\Xi^{0} \\
    \Xi^{-}  \\
  \end{pmatrix}
  $.
}.
The effective baryon mass is given by
\begin{equation}
  M_{B}^{\ast}(\sigma,\bm{\delta},\sigma^{\ast})
  = M_{B} - g_{\sigma B}\sigma - g_{\delta B}\bm{\delta}\cdot\bm{I}_{B} - g_{\sigma^{\ast}B}\sigma^{\ast},
\end{equation}
with $M_{B}$ being the free mass and $\bm{I}_{B}$ being the isospin matrix for $B$~\footnote{
  The isospin matrix for $B$ is defined as
  $\bm{I}_{N} = \bm{I}_{\Xi} = \left(\tau_{1},\tau_{2},\tau_{3}\right)$, $\bm{I}_{\Lambda} = \bm{0}$, and $\bm{I}_{\Sigma} = \left(I_{1},I_{2},I_{3}\right)$ \\ with
  $
  \tau_{1} =
  \begin{pmatrix}
    0 & 1 \\
    1 & 0 \\
  \end{pmatrix}
  $,
  $
  \tau_{2} =
  \begin{pmatrix}
    0 & -i \\
    i & 0  \\
  \end{pmatrix}
  $,
  $
  \tau_{3} =
  \begin{pmatrix}
    1 & 0  \\
    0 & -1 \\
  \end{pmatrix}
  $,
  $\displaystyle
  I_{1} = \frac{1}{\sqrt{2}}
  \begin{pmatrix}
    0 & 1 & 0 \\
    1 & 0 & 1 \\
    0 & 1 & 0 \\
  \end{pmatrix}
  $,
  $\displaystyle
  I_{2} = \frac{i}{\sqrt{2}}
  \begin{pmatrix}
    0 & -i & 0 \\
    i & 0 & -i \\
    0 & i & 0  \\
  \end{pmatrix}
  $, and
  $
  I_{3} =
  \begin{pmatrix}
    1 & 0 & 0  \\
    0 & 0 & 0  \\
    0 & 0 & -1 \\
  \end{pmatrix}
  $.
}.
The $\sigma$-$B$, $\omega$-$B$, $\delta$-$B$, $\rho$-$B$, $\sigma^{\ast}$-$B$, and $\phi$-$B$ coupling constants are respectively denoted by $g_{\sigma B}$, $g_{\omega B}$, $g_{\delta B}$, $g_{\rho B}$, $g_{\sigma^{\ast}B}$, and $g_{\phi B}$.
The covariant derivatives for the vector-meson fields are expressed as $W_{\mu\nu}=\partial_{\mu}\omega_{\nu}-\partial_{\nu}\omega_{\mu}$, $\bm{R}_{\mu\nu}=\partial_{\mu}\bm{\rho}_{\nu}-\partial_{\nu}\bm{\rho}_{\mu}$, and $P_{\mu\nu}=\partial_{\mu}\phi_{\nu}-\partial_{\nu}\phi_{\mu}$.
In addition, a nonlinear potential in Eq.~\eqref{eq:Lagrangian} is supplemented as follows:
\begin{align}
  U_{\textrm{NL}}(\sigma,\omega^{\mu},\bm{\rho}^{\mu},\phi^{\mu})
  & = \frac{1}{3}g_{2}\sigma^{3} + \frac{1}{4}g_{3}\sigma^{4}
    - \frac{1}{4}c_{3}\left(\omega_{\mu}\omega^{\mu}\right)^{2}
    - \frac{1}{4}f_{3}\left(\phi_{\mu}\phi^{\mu}\right)^{2}
    \nonumber \\
  & - \Lambda_{\omega\rho}\left(\omega_{\mu}\omega^{\mu}\right)\left(\bm{\rho}_{\nu}\cdot\bm{\rho}^{\nu}\right)
    - \Lambda_{\phi\rho}\left(\phi_{\mu}\phi^{\mu}\right)\left(\bm{\rho}_{\nu}\cdot\bm{\rho}^{\nu}\right).
    \label{eq:NL}
\end{align}
The first and second terms in Eq.~\eqref{eq:NL} are introduced to provide a quantitative description of the ground-state properties of symmetric nuclear matter~\citep{Boguta:1977xi,Lalazissis:1996rd}.
We introduce the quartic $\omega$  self-interaction and the $\omega$-$\rho$ mixing, which only affects the characteristics of $N\not=Z$ finite nuclei and isospin-asymmetric nuclear matter~\citep{Sugahara:1993wz,Mueller:1996pm,Todd-Rutel:2005yzo,Pradhan:2022txg,Malik:2024qjw}.
Furthermore, the quartic self-interaction of the $\phi$ meson and the $\phi$-$\rho$ mixing are taken into account to sustain the model parameters determined in SU(6) symmetry when the vector fields are extended to the SU(3) flavor-symmetry scheme~\citep{Wadhwa:2025mae}.

In the present study, the hadron and lepton masses in free space are taken as follows: $M_{N}=939$ MeV, $M_{\Lambda}=1116$ MeV, $M_{\Sigma}=1193$ MeV, $M_{\Xi}=1318$ MeV, $m_{\sigma}=465$ MeV, $m_{\omega}=783$ MeV, $m_{\delta}=980$ MeV, $m_{\rho}=775$ MeV, $m_{\sigma^{\ast}}=990$ MeV, $m_{\phi}=1019$ MeV, $m_{e}=0.511$ MeV, and $m_{\mu}=106$ MeV.
\RevA{Note that the $\sigma$-meson mass, $m_{\sigma}$, is adjusted to reproduce the ground-state properties of several closed-shell nuclei (see Sec.~\ref{sec:model}).}

In mean-field approximation, the meson fields are replaced by the mean-field values: $\bar{\sigma}$, $\bar{\omega}$, $\bar{\delta}$, $\bar{\rho}$, $\bar{\sigma}^{\ast}$, and $\bar{\phi}$.
The equations of motion for the meson fields in uniform matter are, thus, given by
\begin{align}
  m_{\sigma}^{2}\bar{\sigma} + g_{2}\bar{\sigma}^{2} + g_{3}\bar{\sigma}^{3}
  & = \sum_{B}g_{\sigma B}\rho_{B}^{s},
    \label{eq:sigma} \\
  m_{\omega}^{2}\bar{\omega} + c_{3}\bar{\omega}^{3} + 2\Lambda_{\omega\rho}\bar{\omega}\bar{\rho}^{2}
  & = \sum_{B}g_{\omega B}\rho_{B},
    \label{eq:omega} \\
  m_{\delta}^{2}\bar{\delta}
  & = \sum_{B}g_{\delta B}\rho_{B}^{s}\left(\bm{I}_{B}\right)_{3},
    \label{eq:delta} \\
  m_{\rho}^{2}\bar{\rho} + 2\Lambda_{\omega\rho}\bar{\omega}^{2}\bar{\rho} + 2\Lambda_{\phi\rho}\bar{\phi}^{2}\bar{\rho}
  & = \sum_{B}g_{\rho B}\rho_{B}\left(\bm{I}_{B}\right)_{3},
    \label{eq:rho} \\
  m_{\sigma^{\ast}}^{2}\bar{\sigma}^{\ast}
  & = \sum_{B}g_{\sigma^{\ast}B}\rho_{B}^{s},
    \label{eq:sigma-star} \\
  m_{\phi}^{2}\bar{\phi} + f_{3}\bar{\phi}^{3} + 2\Lambda_{\phi\rho}\bar{\phi}\bar{\rho}^{2}
  & = \sum_{B}g_{\phi B}\rho_{B},
    \label{eq:phi}
\end{align}
where the scalar density, $\rho_{B}^{s}$, and the baryon density, $\rho_{B}$, read
\begin{align}
  \rho_{B}^{s}
  & = \frac{1}{\pi^{2}}\int_{0}^{k_{F_{B}}}dk \, k^{2} \frac{M_{B}^{\ast}}{\sqrt{k^{2}+M_{B}^{\ast2}}}, \\
  \rho_{B}
  & = \frac{k_{F_{B}}^{3}}{3\pi^{2}},
\end{align}
with $k_{F_{B}}$ being the Fermi momentum for $B\in\{p,\,n,\,\Lambda,\,\Sigma^{+},\,\Sigma^{0},\,\Sigma^{-},\,\Xi^{0},\,\Xi^{-}\}$.
The total energy density, $\varepsilon$, and pressure, $P$, in neutron-star matter are expressed as
\begin{align}
  \varepsilon
  & = \sum_{B}\varepsilon_{B} + \varepsilon_{M} + \sum_{\ell}\varepsilon_{\ell}, \\
  P
  & = \sum_{B}P_{B} + P_{M} + \sum_{\ell}P_{\ell},
\end{align}
where the baryon ($B$), meson ($M$), and lepton ($\ell$) contributions to $\varepsilon$ and $P$ are given by
\begin{align}
  \varepsilon_{B}
  & = \frac{1}{\pi^{2}} \int_{0}^{k_{F_{B}}}dk \, k^{2} \sqrt{k^{2}+M_{B}^{\ast2}}, \\
  \varepsilon_{M}
  & = \frac{1}{2} \left(m_{\sigma}^{2}\bar{\sigma}^{2} + m_{\omega}^{2}\bar{\omega}^{2} + m_{\delta}^{2}\bar{\delta}^{2}
    + m_{\rho}^{2}\bar{\rho}^{2} + m_{\sigma^{\ast}}^{2}\bar{\sigma}^{\ast2} + m_{\phi}^{2}\bar{\phi}^{2} \right)
    \nonumber \\
  & + \frac{1}{3}g_{2}\bar{\sigma}^{3} + \frac{1}{4}g_{3}\bar{\sigma}^{4} + \frac{3}{4}c_{3}\bar{\omega}^{4} + \frac{3}{4}f_{3}\bar{\phi}^{4}
    + 3\Lambda_{\omega\rho}\bar{\omega}^{2}\bar{\rho}^{2} + 3\Lambda_{\phi\rho}\bar{\phi}^{2}\bar{\rho}^{2}, \\
  \varepsilon_{\ell}
  & = \frac{1}{\pi^{2}} \int_{0}^{k_{F_{\ell}}}dk \, k^{2} \sqrt{k^{2}+m_{\ell}^{2}},
\end{align}
and
\begin{align}
  P_{B}
  & = \frac{1}{3\pi^{2}} \int_{0}^{k_{F_{B}}}dk \, \frac{k^{4}}{\sqrt{k^{2}+M_{B}^{\ast2}}}, \\
  P_{M}
  & = -\frac{1}{2} \left(m_{\sigma}^{2}\bar{\sigma}^{2} - m_{\omega}^{2}\bar{\omega}^{2} + m_{\delta}^{2}\bar{\delta}^{2}
    - m_{\rho}^{2}\bar{\rho}^{2} + m_{\sigma^{\ast}}^{2}\bar{\sigma}^{\ast2} - m_{\phi}^{2}\bar{\phi}^{2} \right)
    \nonumber \\
  & - \frac{1}{3}g_{2}\bar{\sigma}^{3} - \frac{1}{4}g_{3}\bar{\sigma}^{4} + \frac{1}{4}c_{3}\bar{\omega}^{4} + \frac{1}{4}f_{3}\bar{\phi}^{4}
    + \Lambda_{\omega\rho}\bar{\omega}^{2}\bar{\rho}^{2} + \Lambda_{\phi\rho}\bar{\phi}^{2}\bar{\rho}^{2}, \\
  P_{\ell}
  & = \frac{1}{3\pi^{2}} \int_{0}^{k_{F_{\ell}}}dk \, \frac{k^{4}}{\sqrt{k^{2}+m_{\ell}^{2}}}.
\end{align}

\section{SU(3) symmetry in the vector-meson couplings}
\label{sec:SU3}

To study the properties of neutron stars with hyperons ($Y$), it is important to extend SU(6) spin-flavor symmetry based on the non-relativistic quark model to the more general SU(3) flavor symmetry~\citep{Weissenborn:2011ut,Miyatsu:2011bc,Katayama:2012ge,Sulaksono:2012ny,Miyatsu:2013yta,Lopes:2013cpa,Oertel:2014qza,Miyatsu:2015kwa,Li:2018jvz,Spinella:2018dab,Fu:2022eeb,Lopes:2022vjx}.
Restricting the discussion to three quark flavors (up, down, and strange), SU(3) symmetry can be regarded as a symmetry group of strong interaction.
To consider combinations of the meson-baryon couplings, it is convenient to employ the SU(3)-invariant interaction Lagrangian.
Using the matrix representations for the baryon octet, $B$, and meson nonet (singlet state, $M_{1}$, and octet state, $M_{8}$), the interaction Lagrangian can be written as the sum of three terms: one arising from the coupling of the meson singlet to the baryon octet ($S$ term) and the other two terms from the interaction of the meson octet and the baryons---one being the antisymmetric ($F$) term and the other being the symmetric ($D$) term~\citep{deSwart:1963pdg,Lichtenberg:1978pc}:
\begin{equation}
  \mathcal{L}_{\textrm{int}} = - g_{8} \sqrt{2} \left[ \alpha \textrm{Tr}\left(\left[\bar{B},M_{8}\right] B\right)
    + (1-\alpha) \textrm{Tr}\left(\left\{\bar{B},M_{8}\right\} B\right) \right]
  - g_{1} \frac{1}{\sqrt{3}} \textrm{Tr}\left(\bar{B}B\right)\textrm{Tr}\left(M_{1}\right),
  \label{eq:SU3-Lagrangian}
\end{equation}
where $g_{1}$ and $g_{8}$ are respectively the coupling constants for the meson singlet and octet states, and $\alpha$ ($0 \leq \alpha \leq 1$) is known as the $F/(F+D)$ ratio.

Here we focus on the vector-meson couplings to the octet baryons, because, as usual, the other coupling constants can be determined so as to reproduce the observed properties of nuclear matter and hypernuclei~\footnote{
  When SU(3) symmetry is applied to the {\it isovector}, vector mesons, the Fock term is, in fact, necessary to reproduce the observed symmetry energy~\citep{Katayama:2012ge}. }.
The physical $\omega$ and $\phi$ mesons are described in terms of the pure singlet, $\ket{1}$, and octet, $\ket{8}$, states as
\begin{align}
  \omega
  &= \cos \theta_{v} \ket{1} + \sin \theta_{v} \ket{8}, \\
  \phi
  &= - \sin \theta_{v} \ket{1} + \cos \theta_{v} \ket{8},
    \label{eq:mixing}
\end{align}
with $\theta_{v}$ being the mixing angle for the vector mesons~\footnote{
  The matrix representations for $M_{1}$, $M_{8}$, and $B$ are expressed as 
  $\displaystyle M_{1}=\frac{1}{\sqrt{3}}\textrm{diag}\left(\phi_{1},\phi_{1},\phi_{1}\right)$, \\
  $
  M_{8} =
  \begin{pmatrix}
    \rho^{0}/\sqrt{2}+\omega_{8}/\sqrt{6} & \rho^{+} & K^{\ast+}  \\
    \rho^{-} & -\rho^{0}/\sqrt{2}+\omega_{8}/\sqrt{6} & K^{\ast0} \\
    K^{\ast-} & \bar{K}^{\ast0} & -2\omega_{8}/\sqrt{6}           \\
  \end{pmatrix}
  $, and 
  $
  B =
  \begin{pmatrix}
    \Sigma^{0}/\sqrt{2}+\Lambda/\sqrt{6} & \Sigma^{+} & p  \\
    \Sigma^{-} & -\Sigma^{0}/\sqrt{2}+\Lambda/\sqrt{6} & n \\
    \Xi^{-} & \Xi^{0} & -2\Lambda/\sqrt{6}                 \\
  \end{pmatrix}
  $ with $\phi_{1}$ ($\omega_{8}$) being the singlet (octet) state.
}.

In SU(3) symmetry, all possible combinations of the couplings are then determined by four parameters: the singlet and octet coupling constants, $g_{1}$ and $g_{8}$, the $F/(F+D)$ ratio for the vector mesons, $\alpha_{v}$, and the mixing angle, $\theta_{v}$.
With the coupling ratio defined by $z_{v}=g_{8}/g_{1}$, the relations of the coupling constants for the $\omega$ and $\phi$ mesons in SU(3) symmetry can be expressed as
\begin{align}
  g_{\omega\Lambda}
  &= \frac{1-\frac{2}{\sqrt{3}}z_{v}(1-\alpha_{v})\tan\theta_{v}}{1-\frac{1}{\sqrt{3}}z_{v}(1-4\alpha_{v})\tan\theta_{v}} g_{\omega N},
  \label{eq:g-omega-Lambda} \\
  g_{\omega\Sigma}
  &= \frac{1+\frac{2}{\sqrt{3}}z_{v}(1-\alpha_{v})\tan\theta_{v}}{1-\frac{1}{\sqrt{3}}z_{v}(1-4\alpha_{v})\tan\theta_{v}} g_{\omega N}, \\
  g_{\omega\Xi}
  &= \frac{1-\frac{1}{\sqrt{3}}z_{v}(1+2\alpha_{v})\tan\theta_{v}}{1-\frac{1}{\sqrt{3}}z_{v}(1-4\alpha_{v})\tan\theta_{v}} g_{\omega N}, \\
  g_{\phi N}
  &= - \frac{\tan\theta_{v}+\frac{1}{\sqrt{3}}z_{v}(1-4\alpha_{v})}{1-\frac{1}{\sqrt{3}}z_{v}(1-4\alpha_{v})\tan\theta_{v}} g_{\omega N},
  \label{eq:g-phi-N} \\
  g_{\phi\Lambda}
  &= - \frac{\tan\theta_{v}+\frac{2}{\sqrt{3}}z_{v}(1-\alpha_{v})}{1-\frac{1}{\sqrt{3}}z_{v}(1-4\alpha_{v})\tan\theta_{v}} g_{\omega N}, \\
  g_{\phi\Sigma}
  &= - \frac{\tan\theta_{v}-\frac{2}{\sqrt{3}}z_{v}(1-\alpha_{v})}{1-\frac{1}{\sqrt{3}}z_{v}(1-4\alpha_{v})\tan\theta_{v}} g_{\omega N}, \\
  g_{\phi\Xi}
  &= - \frac{\tan\theta_{v}+\frac{1}{\sqrt{3}}z_{v}(1+2\alpha_{v})}{1-\frac{1}{\sqrt{3}}z_{v}(1-4\alpha_{v})\tan\theta_{v}} g_{\omega N}.
\end{align}
The $\rho$-$Y$ coupling constants are given by
\begin{equation}
  g_{\rho\Lambda} = 0, \quad
  g_{\rho\Sigma} = 2\alpha_{v}g_{\rho N}, \quad
  g_{\rho\Xi} = -(1-2\alpha_{v})g_{\rho N}.
  \label{eq:g-rho-Y}
\end{equation}
Conventionally, vector couplings are taken with $\alpha_{v}=1$, under which the contribution from the $D$ term is ignored.
In the present study, by relaxing this assumption and treating $\alpha_{v}$ as a free parameter, we are able to examine the impact of both the $F$ and $D$ terms.
Although an additional interaction between $\Lambda$ and $\Sigma^{0}$ via the $\rho$ meson should be treated in SU(3) symmetry under the relation, $g_{\rho\Lambda\Sigma}=\frac{2}{\sqrt{3}}(1-\alpha_{v})g_{\rho N}$~\citep{Lopes:2023gzj}, we neglect its effect in the present study since our interest lies in the $NN$ and $NY$ interactions.

\RevD{We summarize again the physical meanings of the parameters in SU(3) flavor symmetry. Specifically, $\alpha_{v}$ represents the $F/(F+D)$ ratio, which controls the relative strength of the antisymmetric and symmetric couplings; $\theta_{v}$ is the singlet-octet mixing angle that determines the physical composition of the $\omega$ and $\phi$ mesons; and $z_{v}$ denotes the singlet-to-octet coupling ratio, which governs the overall magnitude of the vector-meson couplings.}

\section{Model parameters}
\label{sec:model}

\subsection{Model construction and parameters for nucleons}

Here, we construct the effective interaction that accounts not only for the characteristics of finite nuclei but also for the properties of dense nuclear matter.
As a first step, the model optimization is performed so as to fit the experimental data for binding energy per nucleon, $B/A$, and charge radius, $R_{\textrm{ch}}$, of several, closed-shell nuclei in SU(6) symmetry~\citep{Sugahara:1993wz,Miyatsu:2023lki,Miyatsu:2024ioc}.
The characteristics of several finite nuclei and the coupling constants for $N$ are listed in Tables~\ref{tab:nuclei} and \ref{tab:coupling-nucleon}.
The bulk properties of nuclear matter at the saturation density, $n_{0}=0.15$ fm$^{-3}$, are calculated as follows: the effective nucleon mass $M_{N}^{\ast}/M_{N}=0.70$, binding energy per nucleon $E_{0}(n_{0})=-16.35$ MeV, nuclear incompressibility $K_{0}=240$ MeV, nuclear symmetry energy $E_{\textrm{sym}}(n_{0})=32.5$ MeV, slope parameter $L=50$ MeV, and curvature parameter $K_{\textrm{sym}}=-210.67$ MeV.
We emphasize that the $\delta$-$N$ coupling constant, $g_{\delta N}$, is determined to reproduce the proton-neutron effective mass splitting, $M_{p}^{\ast}-M_{n}^{\ast}\simeq100$ MeV, at $n_{0}$~\citep{vanDalen:2006pr,Miyatsu:2022wuy}.
Furthermore, the self-interaction of the $\omega$ meson is introduced to satisfy the constraints from elliptical flow data and kaon production data in heavy-ion collisions\RevC{, particularly the pressure range of symmetric nuclear matter at high densities}~\citep{Danielewicz:2002pu,Fuchs:2005zg,Lynch:2009vc}.

As a second step, we determine the coupling constants for $N$ in SU(3) flavor symmetry.
According to Eq.~\eqref{eq:g-phi-N}, the $\phi$ meson, in addition to the $\omega$ meson, influences the nuclear EoS in SU(3) symmetry.
To preserve the same saturation properties---$E_{0}(n_{0})$, $K_{0}$, $E_{\textrm{sym}}(n_{0})$, $L$, and $K_{\textrm{sym}}$---as in the SU(6) framework, we adjust the vector-meson coupling constants, $g_{\omega N}$, $g_{\phi N}$, $c_{3}$ and $\Lambda_{\omega\rho}$, by introducing two additional couplings, $f_{3}$ and $\Lambda_{\phi\rho}$ (see Eq.~\eqref{eq:NL}).
Several parameter sets in SU(3) symmetry are also listed in Table~\ref{tab:coupling-nucleon}.
\RevA{When the quartic $\phi$-meson self-interaction and $\phi$-$\rho$ mixing terms are omitted,}
\RevD{the extension to SU(3) flavor symmetry slightly disturbs the original SU(6) saturation conditions, particularly $K_{0}$ and $L$~\citep{Miyatsu:2013yta}.}

\begin{table}[t!]
  \caption{\label{tab:nuclei}
    Theoretical predictions for ground-state properties of several closed-shell nuclei in SU(6) symmetry.
    Experimental data for the binding energy per nucleon, $B/A$, and charge radius, $R_{\rm ch}$, are referred to \citet{Wang:2021xhn} and \citet{Angeli:2013epw}, respectively.
    The neutron skin thickness, $R_{\rm skin}$, is defined as the difference between the root-mean-square radii of point neutrons and protons in a nucleus.
    As explained in Refs.~\citep{Miyatsu:2023lki,Miyatsu:2024ioc}, it is difficult to understand the PREX-2 and CREX results simultaneously in the present study~\citep{PREX:2021umo,CREX:2022kgg}.
  }
  \begin{ruledtabular}
    \begin{tabular}{lccccr}
      \multirow{2}{*}{Nucleus} & \multicolumn{2}{c}{$B/A$ (MeV)} & \multicolumn{2}{c}{$R_{\rm ch}$ (fm)} & \multirow{2}{*}{$R_{\rm skin}$ (fm)} \\
      \cline{2-3}\cline{4-5}
      \          & Theory & Exp. & Theory & Exp.    &       \ \\
      \colrule
      $^{16}$O   &   7.98 & 7.98 &   2.74 & 2.70 & $-0.03$ \\
      $^{40}$Ca  &   8.55 & 8.55 &   3.47 & 3.48 & $-0.05$ \\
      $^{48}$Ca  &   8.48 & 8.67 &   3.50 & 3.48 &   0.20  \\
      $^{68}$Ni  &   8.61 & 8.68 &   3.90 & 3.89 &   0.21  \\
      $^{90}$Zr  &   8.62 & 8.71 &   4.29 & 4.27 &   0.09  \\
      $^{132}$Sn &   8.37 & 8.35 &   4.74 & 4.71 &   0.28  \\
      $^{208}$Pb &   7.86 & 7.87 &   5.54 & 5.50 &   0.21  \\
    \end{tabular}
  \end{ruledtabular}
\end{table}

\begin{table*}[t!]
  \caption{\label{tab:coupling-nucleon}
    Coupling constants for $N$ in SU(6) and SU(3) symmetry.
    \RevA{Since the $\sigma^{\ast}$ meson is mainly composed of an $s\bar{s}$ pair, w}e assume that the $\sigma^{\ast}$ meson does not couple to $N$ due to the OZI rule, namely $g_{\sigma^{\ast}N}=0$~\citep{Schaffner:1995th}.
    In the limit of the \textit{ideal} mixing, the $F/(F+D)$ ratio, mixing angle, and singlet-to-octet coupling ratio are respectively given by $\alpha_{v}^{id}=1.00$, $\theta_{v}^{id}=\tan^{-1}(1/\sqrt{2})\simeq35.26^{\circ}$, and $z_{v}^{id}=1/\sqrt{6}\simeq0.4082$.
    In SU(3) symmetry, we present the cases for $\theta_{v}=\theta_{v}^{id}$ from the \textit{ideal} mixing (case A) and $\theta_{v}=36.5^{\circ}$ from \RevA{Particle Data Group (PDG)} (case B)~\citep{ParticleDataGroup:2024cfk}.
    The parameter $g_{2}$ is in fm$^{-1}$.
  }
  \begin{ruledtabular}
    \begin{tabular}{lccccccccccccc}
      Sym.                   & Case               & $(\alpha_{v},\theta_{v},z_{v})$
      & $g_{\sigma N}$ & $ g_{\omega N}$ & $g_{\delta N}$ & $g_{\rho N}$ & $g_{\phi N}$ & $g_{2}$ &  $g_{3}$ & $c_{3}$ & $f_{3}$ & $\Lambda_{\omega\rho}$ & $\Lambda_{\phi\rho}$ \\
      \colrule
      SU(6)                  & \                  & $(\alpha_{v}^{id},\theta_{v}^{id},z_{v}^{id})$
      &           8.36 &           10.77 &           6.71 &         7.45 &          --- &   11.89 & $-18.91$ &   10.00 &     --- &                 241.06 &                  --- \\
      \colrule
      \multirow{6}{*}{SU(3)} & \multirow{3}{*}{A} & $(\alpha_{v}^{id},\theta_{v}^{id},0.00)$
      &           8.36 &            9.46 &           6.71 &         7.45 &      $-6.69$ &   11.89 & $-18.91$ &   12.97 &  125.85 &                 241.10 &                408.76 \\
      \                      & \                  & $(\alpha_{v}^{id},\theta_{v}^{id},0.25)$
      &           8.36 &           10.63 &           6.71 &         7.45 &      $-2.23$ &   11.89 & $-18.91$ &   10.27 & 1096.49 &                 241.07 &                408.02 \\
      \                      & \                  & $(\alpha_{v}^{id},\theta_{v}^{id},0.50)$
      &           8.36 &           10.74 &           6.71 &         7.45 &         1.06 &   11.89 & $-18.91$ &   10.06 & 4637.76 &                 241.06 &                407.25 \\
      \cline{2-14}
      \                      & \multirow{3}{*}{B} & $(0.25,36.5^{\circ},0.15)$
      &           8.36 &            9.36 &           6.71 &         7.45 &      $-6.93$ &   11.89 & $-18.91$ &   13.24 &  117.40 &                 241.10 &                408.76 \\
      \                      & \                  & $(0.50,36.5^{\circ},0.15)$
      &           8.36 &            9.74 &           6.71 &         7.45 &      $-5.98$ &   11.89 & $-18.91$ &   12.25 &  156.82 &                 241.10 &                408.66 \\
      \                      & \                  & $(0.75,36.5^{\circ},0.15)$
      &           8.36 &           10.05 &           6.71 &         7.45 &      $-5.05$ &   11.89 & $-18.91$ &   11.50 &  220.33 &                 241.09 &                408.61 \\
    \end{tabular}
  \end{ruledtabular}
\end{table*}

\subsection{Coupling constants for hyperons}

\begin{table*}[t!]
  \caption{\label{tab:coupling-hyperon}
    Coupling constants for $Y$ in SU(6) and SU(3) symmetry.
    We assume that $g_{\delta\Lambda}=0$, $g_{\delta\Sigma}=2g_{\delta N}$, and $g_{\sigma^{\ast}\Sigma}=g_{\sigma^{\ast}\Lambda}$.
    For detail, see the text.
  }
  \begin{ruledtabular}
    \begin{tabular}{lcccccccc}
      Sym.                   & Case               & $(\alpha_{v},\theta_{v},z_{v})$
      & $g_{\sigma\Lambda}$ & $ g_{\sigma\Sigma}$ & $g_{\sigma\Xi}$ & $g_{\delta\Xi}$ & $g_{\sigma^{\ast}\Lambda}$ & $g_{\sigma^{\ast}\Xi}$ \\
      \colrule
      SU(6)                  & \                  & $(\alpha_{v}^{id},\theta_{v}^{id},z_{v}^{id})$
      &                5.12 &                3.41 &            2.77 &            7.07 &                    $-3.74$ &                $-9.99$ \\
      \colrule
      \multirow{6}{*}{SU(3)} & \multirow{3}{*}{A} & $(\alpha_{v}^{id},\theta_{v}^{id},0.00)$
      &                7.27 &                5.55 &            7.06 &            5.88 &                       ---~\footnote{Because the $\sigma$ meson contribution already gives $U_{\Lambda}^{(\Lambda)}(n_{0}/2)\le-5$ MeV, the additional attractive force due to the $\sigma^{\ast}$ meson is not required.}  & ---~$^{\textrm{a}}$ \\
      \                      & \                  & $(\alpha_{v}^{id},\theta_{v}^{id},0.25)$
      &                6.05 &                4.34 &            4.64 &            6.55 &                    $-2.37$ &                $-7.50$ \\
      \                      &\                   & $(\alpha_{v}^{id},\theta_{v}^{id},0.50)$
      &                4.64 &                2.92 &            1.80 &            7.45 &                    $-4.02$ &                $-9.29$ \\
      \cline{2-9}
      \                      & \multirow{3}{*}{B} & $(0.25,36.5^{\circ},0.15)$
      &                7.07 &                5.75 &            6.87 &         $-1.09$ &        ---~$^{\textrm{a}}$ &    ---~$^{\textrm{a}}$ \\
      \                      & \                  & $(0.50,36.5^{\circ},0.15)$
      &                6.94 &                5.55 &            6.58 &         $-5.20$ &        ---~$^{\textrm{a}}$ &    ---~$^{\textrm{a}}$ \\
      \                      & \                  & $(0.75,36.5^{\circ},0.15)$
      &                6.79 &                5.27 &            6.20 &            5.13 &        ---~$^{\textrm{a}}$ &    ---~$^{\textrm{a}}$ \\
    \end{tabular}
  \end{ruledtabular}
\end{table*}

Using Eqs.~\eqref{eq:g-omega-Lambda}--\eqref{eq:g-rho-Y}, the hyperon coupling constants for the vector mesons, $g_{\omega Y}$, $g_{\rho Y}$, and $g_{\phi Y}$, are automatically determined by $\alpha_{v}$, $\theta_{v}$, and $z_{v}$.
On the other hand, those for the scalar mesons are fixed so as to satisfy the empirical data on potential depths.
We here adopt the single-baryon potential based on the so-called Schr\"{o}dinger-equivalent potential~\citep{Jaminon:1981xg,Chen:2007ih}:
\begin{equation}
  U_{B} = \Sigma_{B}^{s} - \Sigma_{B}^{0} + \frac{1}{2M_{B}}\left(\Sigma_{B}^{s}-\Sigma_{B}^{0}\right)^{2},
\end{equation}
where the scalar ($s$) and time ($0$) components of baryon self-energy are written as
\begin{align}
  \Sigma_{B}^{s}
  & = -g_{\sigma B}\bar{\sigma} - g_{\delta B}\bar{\delta}\left(\bm{I}_{B}\right)_{3} - g_{\sigma^{\ast}B}\bar{\sigma}^{\ast}, \\
  \Sigma_{B}^{0}
  & = -g_{\omega B}\bar{\omega} - g_{\rho B}\bar{\rho}\left(\bm{I}_{B}\right)_{3} - g_{\phi B}\bar{\phi}.
\end{align}
As for the $\sigma$-$Y$ couplings, we employ the recently updated values of the potential depths in symmetric nuclear matter (SNM), $U_{\Lambda}^{(\textrm{SNM})}(n_{0})=-27.7$ MeV, $U_{\Sigma}^{(\textrm{SNM})}(n_{0})=+30$ MeV, and $U_{\Xi}^{(\textrm{SNM})}(n_{0})=-21$ MeV~\citep{Batty:1997zp,Kohno:2004pb,Friedman:2007zza,Friedman:2022bpw,Friedman:2023ucs,Friedman:2024epf,Friedman:2025sni}.
The $\sigma^{\ast}$-$\Lambda$ and $\sigma^{\ast}$-$\Xi$ coupling constants, $g_{\sigma^{\ast}\Lambda}$ and $g_{\sigma^{\ast}\Xi}$, are calculated using the conventional relations, $U_{\Xi}^{(\Xi)}(n_{0})\simeq2U_{\Lambda}^{(\Lambda)}(n_{0}/2)$, with the $\Lambda\Lambda$ potential from the Nagara event, $U_{\Lambda}^{(\Lambda)}(n_{0}/2)=-5$ MeV~\citep{Schaffner:1993qj,Takahashi:2001nm}.
In addition, we use the potential depth in pure neutron matter (PNM), $U_{\Xi^{-}}^{(\text{PNM})}=+6$ MeV, from the Lattice QCD result by \RevA{Hadrons to Atomic nuclei from Lattice QCD (HAL QCD)} Collaboration to get $g_{\delta\Xi}$ because $\Xi^{-}$ plays an important role in supporting a massive neutron star~\citep{Inoue:2018axd,Inoue:2019jme,Inoue:2021tdt}.
The other couplings are given by the relations based on the quark model: $g_{\delta\Lambda}=0$, $g_{\delta\Sigma}=2g_{\delta N}$, and $g_{\sigma^{\ast}\Sigma}=g_{\sigma^{\ast}\Lambda}$.
The coupling constants for $Y$ in SU(6) and SU(3) symmetry are listed in Table~\ref{tab:coupling-hyperon}.

\RevA{In the present RMF approach, the coupling constants are not arbitrarily chosen but are determined so as to reproduce the characteristics of finite, closed-shell nuclei and saturation properties of nuclear matter.
  Therefore, the uncertainties of these parameters are already strongly constrained by empirical nuclear data.
  Although a quantitative uncertainty analysis is beyond the scope of the present study, we have confirmed that a moderate variation of the key coupling constants does not alter qualitative conclusions regarding the equation of state and neutron star properties.}

\section{Neutron-star properties in SU(3) flavor symmetry}
\label{sec:results}

The charge neutrality and $\beta$ equilibrium conditions are generally imposed in the discussion of neutron-star properties.
Since the neutron-star radii are remarkably sensitive to the low-density EoS which covers the crust region, we employ the realistic EoS for nonuniform matter, in which neutron-rich nuclei and neutron drips out of the nuclei are considered using the Thomas-Fermi calculation with the uniform nuclear EoS based on RHF approximation~\citep{Miyatsu:2013hea}.
As for the crust-core phase transition, we consider the thermo-dynamical method, and it occurs at $n_{B}=0.089$ fm$^{-3}$ in all the cases~\citep{Miyatsu:2024ioc}, where the total baryon density is given by $n_{B}=\sum_{B}\rho_{B}$.

\begin{figure*}[t!]
  \includegraphics[width=16.0cm,keepaspectratio,clip]{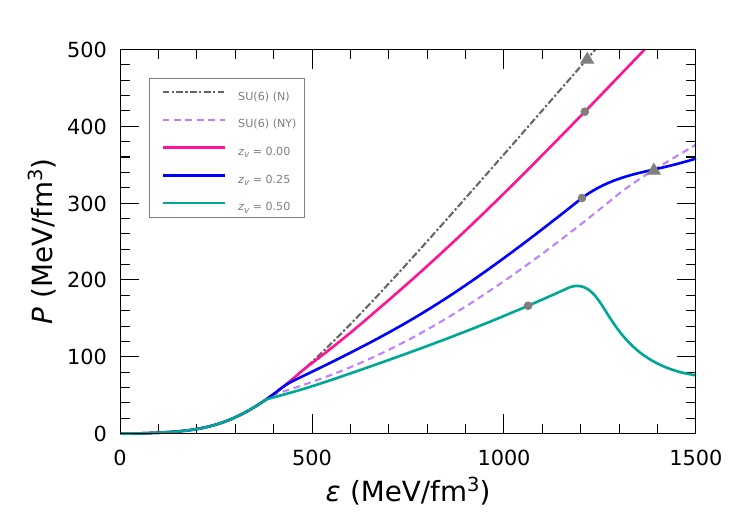}
  \caption{\label{fig:EoS}
    \RevB{Equations of state for neutron stars.
      The SU(6)-symmetry cases are shown with and without hyperons, together with the SU(3)-symmetry cases for $z_{v}=0.00$, $0.25$, and $0.50$ with $\alpha_{v}=\alpha_{v}^{id}$ and $\theta_{v}=\theta_{v}^{id}$ (case A).
      The filled circles and triangles denote the maximum-mass points for SU(3) and SU(6) symmetry, respectively.}
  }
\end{figure*}
\begin{figure*}[t!]
  \includegraphics[width=16.0cm,keepaspectratio,clip]{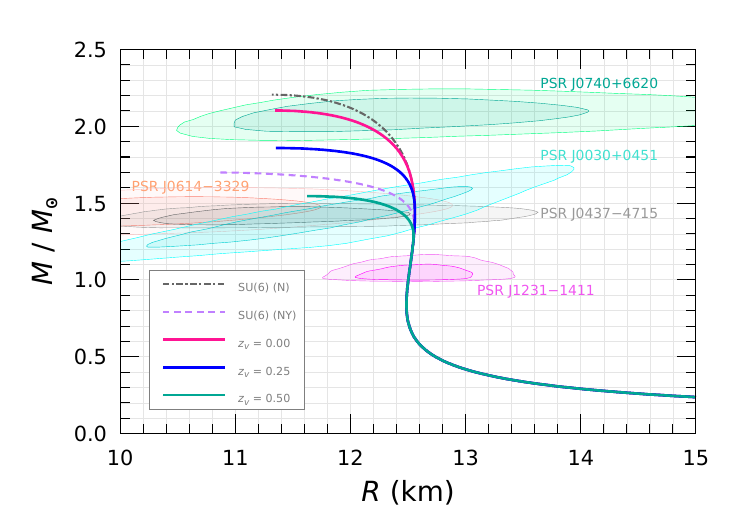}
  \caption{\label{fig:MR}
    Mass-radius relations of neutron stars.
    The observational data are supplemented by the NICER constraints~\citep{Vinciguerra:2023qxq,Choudhury:2024xbk,Salmi:2024aum,Salmi:2024bss,Mauviard:2025dmd}.
    We show the cases for $z_{v}=0.00$, $0.25$, and $0.50$ with $\alpha_{v}=\alpha_{v}^{id}$ and $\theta_{v}=\theta_{v}^{id}$ in SU(3) symmetry (case A).
  }
\end{figure*}
\RevB{The neutron-star EoSs and mass-radius relations of neutron stars are presented in Figs.~\ref{fig:EoS} and \ref{fig:MR}, respectively.}
As is well known, the appearance of hyperons \RevB{softens the EoS and thereby drastically} reduces $M_{\textrm{max}}$ in SU(6) symmetry.
In the SU(3) flavor-symmetry scheme, the smaller $z_{v}$ \RevB{hardens the EoS and} enhances $M_{\textrm{max}}$, and the $2M_{\odot}$ limit can be sufficiently satisfied for $\alpha_{v}=0.00$ ($M_{\textrm{max}}=2.10M_{\odot}$).
Note that as we set $M_{N}^{\ast}/M_{N}=0.70$ at $n_{0}$, the neutron-star radius and dimensionless tidal deformability at the typical mass of a neutron star, $R_{1.4}$ and $\Lambda_{1.4}$, show the relatively small values ($R_{1.4}=12.56$ km and $\Lambda_{1.4}=494$), and thus $\Lambda_{1.4}$ is consistent with the severe constraints from GW170817, $\Lambda_{1.4}=190^{+390}_{-120}$ and $\Lambda_{1.4}=265.18^{+237.88}_{-104.38}$~\citep{LIGOScientific:2018cki,LIGOScientific:2018hze,Huang:2025mrd}.

\begin{figure*}[t!]
  \includegraphics[width=16.0cm,keepaspectratio,clip]{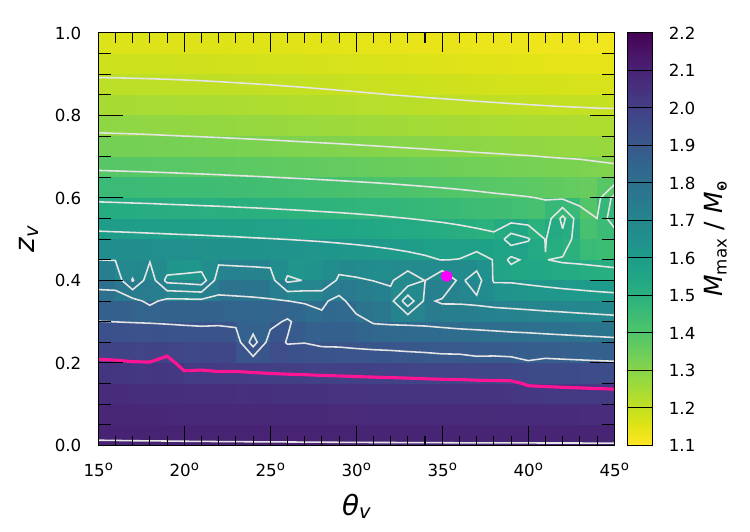}
  \caption{\label{fig:NSmax-alpha}
    Correlation between $\theta_{v}$ and $z_{v}$ in the maximum mass of neutron stars, $M_{\textrm{max}}$, for $\alpha_{v}=\alpha_{v}^{id}$.
    The red thick line is the $2M_{\odot}$ limit, while the white lines denote contours of $M_{\textrm{max}}/M_{\odot}$ at intervals of 0.1.
    The magenta dot represents the case of SU(6) symmetry.
  }
\end{figure*}
The correlation between $\theta_{v}$ and $z_{v}$ in $M_{\textrm{max}}$ is displayed in Fig.~\ref{fig:NSmax-alpha}.
We here fix the \textit{ideal} mixing for $\alpha_{v}$.
It is clearly seen that $\theta_{v}$ shows less impact on $M_{\textrm{max}}$, while $z_{v}$ strongly affects $M_{\textrm{max}}$ and the smaller $z_{v}$ is favorable to support $2M_{\odot}$ neutron stars.
\RevC{When $\alpha_{v}$ is restricted by the realistic mixing angle, $\alpha_{v}^{id}=1.00$, the $2M_{\odot}$ neutron stars are supported for $z_{v}\le0.15$, which is slightly smaller than the result from the Nijmegen extended-soft-core (ESC) model, $z_{v}^{\textrm{ESC}}=0.195$~\citep{Rijken:2010zzb}.}
The irregular contour structures observed around the central band are attributed to the rapid onset of hyperons.
This behavior arises because, near the SU(6) value, the coupling $g_{\phi N}$ becomes small, and in order to reproduce the same saturation properties as in SU(6) symmetry, $f_{3}$ in Eq.~\eqref{eq:NL} must increase significantly (see Table~\ref{tab:coupling-nucleon}).
Consequently, the enhanced contribution of this term leads to the sudden change in hyperon production.

\begin{figure*}[t!]
  \includegraphics[width=16.0cm,keepaspectratio,clip]{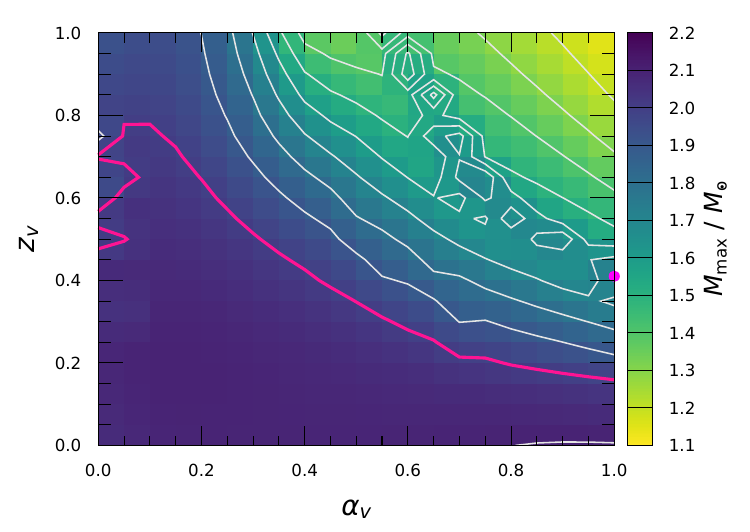}
  \caption{\label{fig:NSmax-theta}
    Correlation between $\alpha_{v}$ and $z_{v}$ in the maximum mass of neutron stars, $M_{\textrm{max}}$, for $\theta_{v}=36.5^{\circ}$~\citep{ParticleDataGroup:2024cfk}.
    The magenta dot represents the case of the \textit{ideal} mixing.
    The colored lines are the same as in Fig.~\ref{fig:NSmax-alpha}.
  }
\end{figure*}
In Fig.~\ref{fig:NSmax-theta}, we also show the correlation between $\alpha_{v}$ and $z_{v}$ in $M_{\textrm{max}}$ with the fixed mixing angle from PDG, $\theta_{v}=36.5^{\circ}$~\citep{ParticleDataGroup:2024cfk}.
Similar to the case of the fixed $\alpha_{v}$ shown in Fig.~\ref{fig:NSmax-alpha}, $z_{v}$ mainly governs the behavior of $M_{\textrm{max}}$.
Nevertheless, $\alpha_{v}$ still plays a qualitative role by setting the stiffness of the neutron-star EoS.
The variation in $\alpha_{v}$ shifts the overall magnitude of $M_{\textrm{max}}$, although it is empirically indicated as $\alpha_{v}=1$.
Thus, $M_{\textrm{max}}$ is primarily driven by $z_{v}$, while $\alpha_{v}$ provides a secondary but non-negligible adjustment to $2M_{\odot}$ neutron stars.

\begin{turnpage}
  \begin{figure*}[p]
    \includegraphics[width=23.0cm,keepaspectratio,clip]{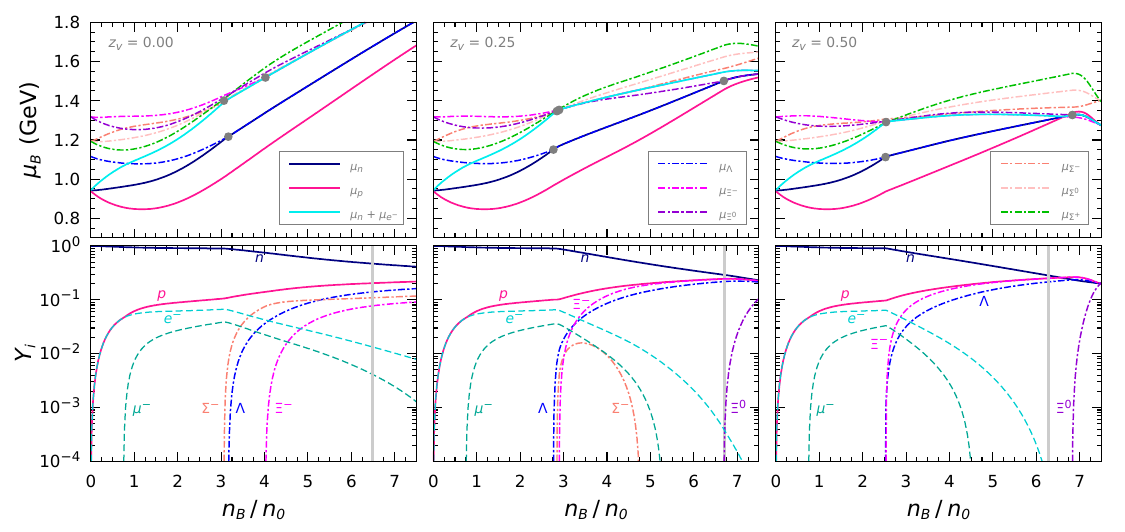}
    \caption{\label{fig:Composition-Z}
      Chemical potential, $\mu_{B}$, and partial fractions, $Y_{i}$, in neutron-star matter as a function of $n_{B}/n_{0}$.
      We show the cases for $z_{v}=0.00$, $0.25$, and $0.50$ with $\alpha_{v}=\alpha_{v}^{id}$ and $\theta_{v}=\theta_{v}^{id}$ in SU(3) symmetry (case A).
      \RevD{The filled circle denotes the onset of hyperons in the upper panels.}
      \RevB{The thick vertical lines in the lower panels shows the density at which a neutron star reaches the maximum-mass point by solving the TOV equation.}
    }
  \end{figure*}
\end{turnpage}

\begin{turnpage}
  \begin{figure*}[p]
    \includegraphics[width=23.0cm,keepaspectratio,clip]{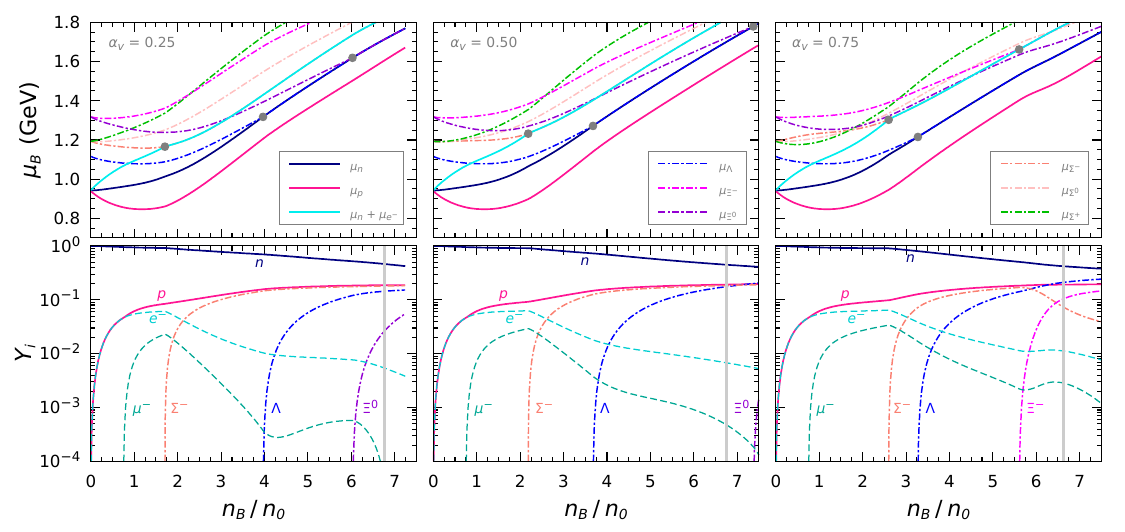}
    \caption{\label{fig:Composition-Alpha}
      Same as in Fig.~\ref{fig:Composition-Z} but for $\alpha_{v}=0.25$, $0.50$, and $0.75$ with $\theta_{v}=36.5^{\circ}$ and $z_{v}=0.15$ (case B).
    }
  \end{figure*}
\end{turnpage}

To investigate the composition of the neutron-star core, the chemical potential, $\mu_{B}=\sqrt{k_{F_{B}}^{2}+M_{B}^{\ast2}}-\Sigma_{B}^{0}$, and partial fractions, $Y_{i}=\rho_{i}/n_{B}$, are presented in Fig.~\ref{fig:Composition-Z}.
As we consider the slightly deeper potentials for $\Xi$ in SNM and PNM, $U_{\Xi}^{(\textrm{SNM})}(n_{0})=-21$ MeV and $U_{\Xi^{-}}^{(\text{PNM})}=+6$ MeV in the present study, $\Xi^{-}$ appears quickly compared with the previous work in \citet{Miyatsu:2013yta}.
In the cases for $z_{v}=0.25$ and $0.50$, $\Lambda$ and $\Xi^{-}$ are generated at the almost same density and their components are enhanced at high densities.
Ultimately, the softening of the neutron-star EoS gives rise to the well-known \textit{hyperon puzzle}.
In contrast, the smaller $z_{v}$ delays the $\Lambda$ onset, and then $\mu_{n}$ continues to increase monotonically at high densities.
Consequently, it enables supporting $2M_{\odot}$ by preventing the rapid hyperon production, although $\Sigma^{-}$ is likely to appear at lower densities through $\mu_{\Sigma^{-}}=\mu_{n}+\mu_{e^{-}}$.

The effect of $\alpha_{v}$ on the hyperon generation is demonstrated in Fig.~\ref{fig:Composition-Alpha}, showing the cases for $\alpha_{v}=0.25$, $0.50$, and $0.75$ with $\theta_{v}=36.5^{\circ}$ and $z_{v}=0.15$.
As already explained in Fig.~\ref{fig:Composition-Z}, the $2M_{\odot}$ constraint is satisfied in all the cases by means of the small $z_{v}$.
As $\alpha_{v}$ decreases, the $\Sigma^{-}$ onset occurs at low densities, and then $\Sigma^{-}$ appears below $2n_{0}$ for $\alpha_{v}=0.25$.
It is interesting that with decreasing $\alpha_{v}$, the chemical potential splitting between $\mu_{\Xi^{0}}$ and $\mu_{\Xi^{-}}$ is enhanced, leading to the situation in which $\Xi^{0}$, rather than $\Xi^{-}$, is produced in the core of a neutron star.

\begin{figure*}[t!]
  \includegraphics[width=16.0cm,keepaspectratio,clip]{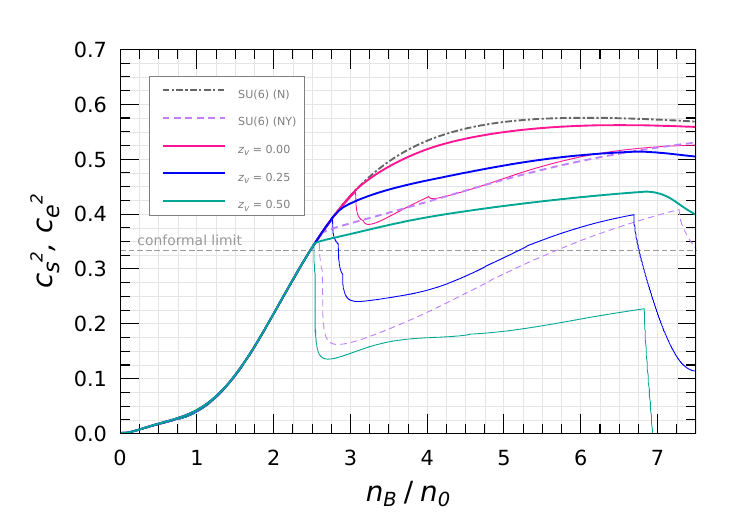}
  \caption{\label{fig:Sound}
    Squared adiabatic and equilibrium speed of sound, $c_{s}^{2}$ (thick) and $c_{e}^{2}$ (thin), in neutron-star matter.
    The gray dashed lines show the conformal limit from QCD~\citep{Bedaque:2014sqa}.
  }
\end{figure*}
The squared speed of sound in neutron-star matter is illustrated in Fig.~\ref{fig:Sound}.
We show two types of speed of sound: one is the adiabatic form, $c_{s}^{2}=\left.\frac{\partial P}{\partial \varepsilon}\right|_{Y_{p}}$, and the other is the equilibrium form, $c_{e}^{2}=\frac{dP}{d\varepsilon}$~\citep{Jaikumar:2021jbw,Aguirre:2022zxn,Tran:2022dva}.
Taking hyperons into account, the difference between $c_{s}^{2}$ and $c_{e}^{2}$ becomes more pronounced.
As explained in \citet{Motta:2020xsg} and \citet{Ye:2024meg}, an abrupt decrease appears around $3n_{0}$, which corresponds to the threshold for the first hyperon (see Fig.~\ref{fig:Composition-Z}).
Moreover, for $z_{v}\ge0.25$, $c_{e}^{2}$ exhibits a sharp decline at high densities, due to the rapid emergence of $\Xi^{0}$.

\section{Summary and conclusion}
\label{sec:summary}

Under SU(3) flavor symmetry, we have studied the properties of neutron-star matter with hyperons using the RMF model.
The model parameters are calibrated to reproduce finite-nucleus observables as well as bulk properties of dense nuclear matter.
To simultaneously satisfy the astrophysical constraints on $\Lambda_{1.4}$ from GW170817 and the experimental data on the dense nuclear EoS from heavy-ion collisions, we have fixed a relatively large $M_{N}^{\ast}$ at $n_{0}$, $M_{N}^{\ast}/M_{N}=0.70$, and have included a quartic self-interaction of the $\omega$ meson.
The $\delta$ meson is incorporated to reproduce the proton-neutron effective mass splitting suggested by DBHF calculations.
We further introduce the additional interactions, $\left(\phi_{\mu}\phi^{\mu}\right)^{2}$ and $\left(\phi_{\mu}\phi^{\mu}\right)\left(\bm{\rho}_{\nu}\cdot\bm{\rho}^{\nu}\right)$, so that SU(6)-based couplings can be retained within the SU(3) scheme.
The hyperon-meson couplings are determined by applying SU(3) flavor symmetry to the vector mesons and by constraining the scalar couplings with empirical potential depths in matter.

We have analyzed how $\alpha_{v}$, $\theta_{v}$, and $z_{v}$ affect $M_{\textrm{max}}$ in the SU(3)-invariant vector-meson couplings.
It has been found that $z_{v}$ predominantly controls $M_{\textrm{max}}$, and the $2M_{\odot}$ neutron stars are supported for $z_{v}\le0.15$.
The $\alpha_{v}$ also contributes significantly to sustaining large masses, whereas the variation in $\theta_{v}$ has a minor impact on neutron-star properties.
Within the SU(3) framework, $\Sigma^{-}$ may become as important as, or even more significant than, $\Lambda$ and $\Xi^{-}$, as shown in Fig.~\ref{fig:Composition-Alpha}.
This highlights the need for a more precise determination of $U_{\Sigma}^{\textrm{(SNM)}}$ through detailed analyses of hypernuclei.
Overall, the SU(3)-based approach yields a consistent description of hyperonic neutron-star matter that reconciles nuclear data with astrophysical observations, offering a plausible resolution of the long-standing \textit{hyperon puzzle}.

\section*{Author Contributions}

Conceptualization, T.M.;
methodology, T.M.;
software, T.M.;
investigation, T.M.;
writing---original draft preparation, T.M.;
writing---review and editing, T.M., M.-K.C., K.K. and K.S.;
visualization, T.M.
All authors have read and agreed to the published version of the manuscript.

\section*{Funding}

This work was supported by the Basic Science Research Program through the National Research Foundation of Korea (NRF) under Grant Nos. RS-2025-16066382, RS-2025-16071941, RS-2023-00242196, and RS-2021-NR060129.

\section*{Data Availability Statement}

The experimental and observational data are available in the literature (references included).
The numerical results presented here can be supplied upon request.
Further inquiries can be directed to T.M.

\section*{Conflicts of Interest}

The authors declare no conflicts of interest.

\bibliographystyle{apsrev4-2.bst}
\bibliography{NS-SU3.bib}

\end{document}